\documentclass[conference]{IEEEtran}
\IEEEoverridecommandlockouts
\usepackage{cite}
\usepackage{amsmath,amssymb,amsfonts}
\usepackage{algorithmic}
\usepackage{graphicx}
\usepackage{textcomp}
\usepackage{xcolor}
\def\BibTeX{{\rm B\kern-.05em{\sc i\kern-.025em b}\kern-.08em
    T\kern-.1667em\lower.7ex\hbox{E}\kern-.125emX}}

\usepackage{algorithm}
\usepackage{array}
\usepackage[caption=false,font=normalsize,labelfont=sf,textfont=sf]{subfig}
\usepackage{textcomp}
\usepackage{stfloats}
\usepackage{url}
\usepackage{xcolor}
\usepackage{verbatim}
\usepackage{graphicx}
\usepackage{enumerate}
\usepackage{enumitem}
\usepackage{tabularx,booktabs,textcomp}
\usepackage{amsthm}

\usepackage{amsmath,amssymb,amsthm}

\makeatletter 
\newcommand{\linebreakand}{%
  \end{@IEEEauthorhalign}
  \hfill\mbox{}\par
  \mbox{}\hfill\begin{@IEEEauthorhalign}
}
\makeatother

\begin{document}

\title{A Resilient Power Distribution System using P2P Energy Sharing}

\author{\IEEEauthorblockN{K. Victor Sam Moses Babu}
\IEEEauthorblockA{\textit{Dept. of EEE} \\
\textit{BITS Pilani Hyderabad Campus}\\
Hyderabad, India \\
victor.babu@in.abb.com}
\and
\IEEEauthorblockN{Divyanshi Dwivedi}
\IEEEauthorblockA{\textit{Dept. of Electrical Engineering} \\
\textit{Indian Institute of Technology Hyderabad}\\
Hyderabad, India \\
divyanshi.dwivedi@in.abb.com}
\and
\IEEEauthorblockN{Dr. Pratyush Chakraborty}
\IEEEauthorblockA{\textit{Dept. of EEE} \\
\textit{BITS Pilani Hyderabad Campus}\\
Hyderabad, India \\
pchakraborty@hyderabad.bits-pilani.ac.in}
\and

\linebreakand 

\IEEEauthorblockN{Dr. Pradeep Kumar Yemula}
\IEEEauthorblockA{\textit{Dept. of Electrical Engineering} \\
\textit{Indian Institute of Technology Hyderabad}\\
Hyderabad, India \\
ypradeep@ee.iith.ac.in}
\and
\IEEEauthorblockN{Dr. Mayukha Pal*}
\IEEEauthorblockA{\textit{ABB Ability Innovation Center}\\
\textit{Asea Brown Boveri}\\
Hyderabad, India \\
*mayukha.pal@in.abb.com}
}

\maketitle

\begin{abstract}
The adoption of distributed energy resources (DERs) such as solar panels and wind turbines is transforming the traditional energy grid into a more decentralized system, where microgrids are emerging as a key concept. Peer-to-Peer (P2P) energy sharing in microgrids enhances the efficiency and flexibility of the overall system by allowing the exchange of surplus energy and better management of energy resources. This work analyzes the impact of P2P energy sharing for three cases - within a microgrid, with neighbouring microgrids, and all microgrids combined together in a distribution system. A standard IEEE 123 node test feeder integrated with renewable energy sources is partitioned into microgrids. For P2P energy sharing between microgrids, the results show significant benefits in cost, reduced energy dependence on the grid, and a significant improvement in the system's resilience. We also predicted the energy requirement for a microgrid to evaluate energy resilience for the control and operation of the microgrid. Overall, the analysis provides valuable insights into the performance and sustainability of microgrids with P2P energy sharing.
\end{abstract}


\section{Introduction}
With increasing dependence on electricity and the growing occurrence of extreme events, it has become critical to develop resilient electric power distribution systems \cite{resiliency2}. As stated in a US federal energy regulatory commission report, resilience involves the ability to withstand, respond, adapt, and prevent disruptive events, man-made attacks, and severe technical faults \cite{Nationelectricity}. Unfortunately, as evidenced by the power cuts that affected over 2.5 million consumers in the aftermath of Cyclone Nisarga in 2020 \cite{Cyclone}; natural disasters damage power distribution feeders that affect a significant percentage of the population. Therefore, it is essential to focus on developing and implementing resilient systems to ensure that we are better prepared to handle such events. Several effective solutions have been proposed to increase the electrical distribution system's resilience, including integrating DERs as local and community resources, line strengthening, and microgrid formulation \cite{resiliencyoption}. Several metrics have been proposed on the basis of operational and planning resilience \cite{r3,r2,r4}. These metrics are not capable of appending the entire system into a single observable quantity to provide an efficient evaluation of resilience. Therefore, we propose the percolation threshold (PT) as a metric that considers the entire system and provides an effective assessment of resilience \cite{resiliency}. The PT is a statistical tool that suggests the state transition of the system. Thus PT is used to analyze how end users of a microgrid improve the economic and resilience conditions by engaging in peer-to-peer (P2P) energy sharing \cite{p2p_resiliency1}.

P2P energy sharing is a suitable means for improving the economic situation in a microgrid, allowing consumers to share energy with their peers and exchange surplus energy when required, resulting in cost benefits for all participants \cite{ p2p_resiliency2}. Game theory can be used to analyze peer-to-peer networks in both cooperative and non-cooperative manners \cite{Chakraborty-S, Kalathil}. Several researchers have applied game theory to peer-to-peer energy sharing. Non-cooperative sharing of solar energy under a net-metering policy without time-of-use pricing was explored \cite{Kalathil-S}. A cooperative strategy for sharing energy storage units under net metering with time-of-use pricing \cite{Victor-GM} was proposed. A cooperative model for sharing storage units with cost allocation based on nucleolus that ensures fairness was presented \cite{Yang}. Sharing of individual energy storage units and a combined storage unit for time-of-use prices using coalitional games were analyzed \cite{Chakraborty}.  While these studies provide a cost-benefit analysis, they do not consider energy analysis that takes into account the dependency on the grid. We have presented a method \cite{resiliencycost} through which resilience is quantified through the complex network and visibility graph. A microgrid with the highest energy resilience in a distribution system was considered for the analysis of how peer-to-peer trading improves cost and resilience. But the impact of P2P trading for combined microgrids was not analyzed. 

Thus, in this work, we analyze the impact of DERs and peer-to-peer trading in a microgrid and the effect of sharing among microgrids. Residential houses with solar and energy storage units are considered in the standard IEEE 123 node test feeder. Five microgrids are formed in this distribution system. A P2P network is established between all houses in a microgrid using cooperative game theory, and a model for cost-effective sharing is developed. The visibility graph is employed to compute the PT, and the resilience of the energy network is evaluated for different cases. In the state-of-the-art, energy sharing between microgrids is not evaluated and the prediction of energy resilience in a microgrid is not considered. The developed cost allocation rule provides fair sharing between all the participants of the P2P energy network and also reduces both electricity cost and the energy requirement from the utility grid. This study conducts both energy and cost analysis and uses a quantitative measure to evaluate the resilience of the energy network under different scenarios. 

To further improve the resilience of energy systems, energy prediction may be implemented to manage energy shortages, congestion, blackouts, and other disruptions. There are several methods that could be used for energy prediction, such as deep learning methods like LSTM \cite{LSTM} and ANNs \cite{ANN}, machine learning methods like decision trees and random forests \cite{ML1}, and statistical methods like regression models and autoregressive integrated moving average (ARIMA) \cite{ARIMA,ARIMA1}. We have used ARIMA for energy prediction in microgrids due to its simplicity, flexibility, and ability to predict trends and seasonality in the data.

Key contributions of this work are as follows:
\begin{enumerate}
    \item We partition all possible microgrids in a distribution system and analyze the green energy sharing for three cases: i) within each individual microgrid, ii) neighboring microgrids, and iii) all microgrids.
    \item P2P energy sharing for all houses with DERs is modeled using cooperative game theory. The developed model ensures that all houses rationally participate in energy sharing to receive economic benefits.
    \item We quantify the improvement of resilience due to P2P energy sharing for all cases using a visibility graph by computing the percolation threshold.   
    \item An energy prediction model using the ARIMA algorithm provides information on the energy requirements for a microgrid from the grid and evaluates the energy resilience. 
\end{enumerate}

We present the methods used to measure resilience, the cooperative game model for P2P energy sharing, and energy prediction in Section \ref{section:Methods}. The results of the simulation study with real-world data are discussed in Section \ref{section:Simulation}. Finally, we draw the conclusions in Section \ref{section:Conclusion}.

\section{Materials and Methods}
\label{section:Methods}
\subsection{Resilience measure for Electrical Distribution System}
\label{subsection:Resiliency}

From complex network theory, a mathematical tool named percolation threshold is found suitable for measuring the resilience in an electrical distribution system that captures the phase transition point in any network \cite{resiliency}. It allows a measure of the system's capability to withstand extreme events without interruption in the supply. The PT for resilience evaluation is computed by obtaining the percolation strength of the network using \cite{GNN}:
\begin {equation}
P_\infty (p) = \frac{1}{NE} \cdot \sum_{i=1}^{T} S(p)
\end {equation}
\noindent where $N$ and $E$ are the numbers of nodes and edges of the network respectively, $S(p)$ is the function for bond occupation probability, $p = e/E$; $e$ denotes the removal of edges in the network. $S(p)$ is summed up $T$ independent times and represents the percolation strength of the network. $S(p)$ initially starts with network probability with the smallest cluster of the network. The network’s susceptibility is calculated as \cite{GNN}:
\begin {equation}
 	\chi(p)=  \frac{(1/N^2 E)\sum_{i=1}^{T} (S(p))^2-[P_\infty (p)]^2} {P_\infty (p)} 	 
\end {equation}
Then the significant value of $p$ is considered as the percolation threshold $\rho_m$ where the value of susceptibility is maximum. 
\begin {equation}
    \rho_m=arg{[\max \chi(p)]}
\end {equation}

\begin{figure}[t]
  \centering
  \includegraphics[width=2.9in]{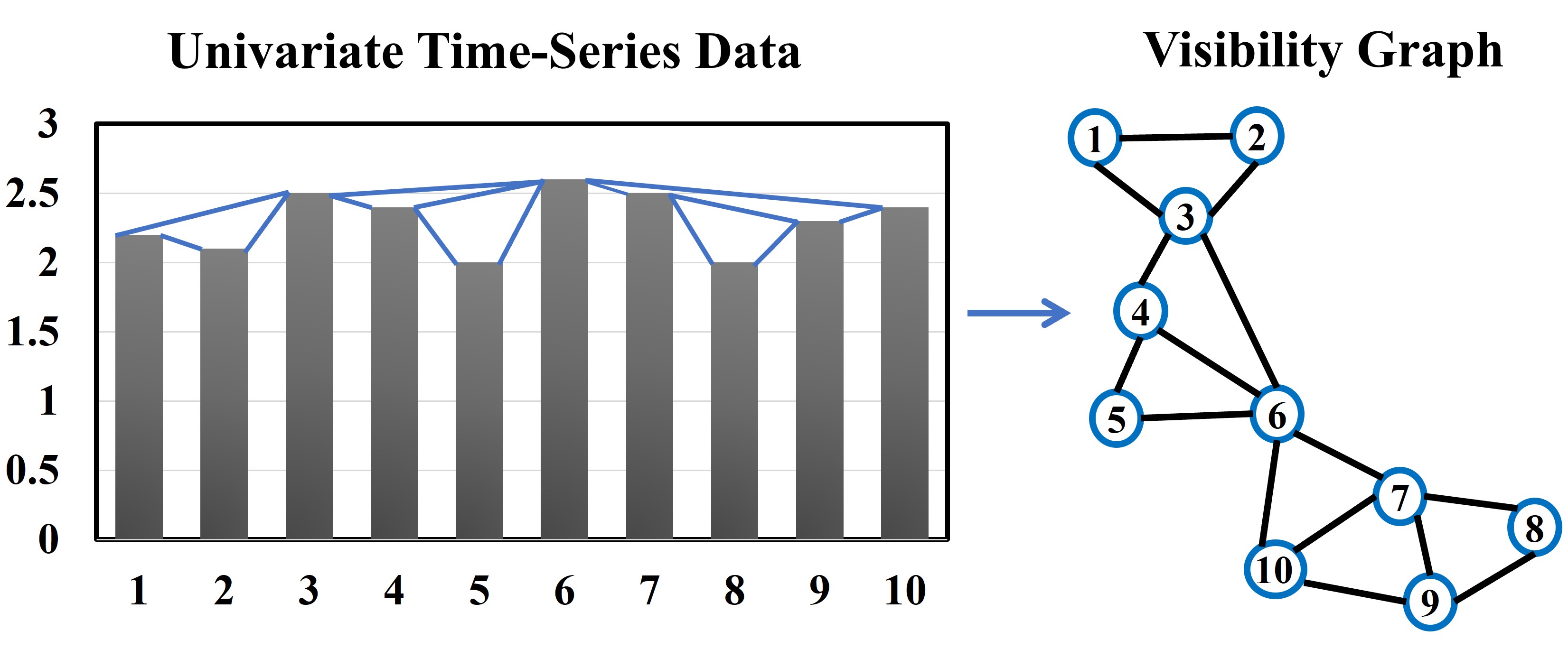}
  \caption{An example for computing complex network from univariate time-
series data using visibility graph.}
  \label{fig:visibility}
\end{figure}

A high value of percolation threshold is expected for the electrical network as it signifies a more resilient network. We need to create a complex network for the distribution system to compute the PT using a visibility graph \cite{Lucas} from the univariate data \cite{resiliencycost}. In consecutive time series, two data points are associated if visibility between the corresponding data and the connection line does not intersect any data height, as shown in Fig. \ref{fig:visibility}. The visibility link is drawn between two data points $(a_m,b_m)$ and $(a_n,b_n)$ if other data point $(a_p,b_p)$ satisfies:
\begin{equation}
    b_p<b_n+(b_m-b_n) \frac{(a_n-a_p)}{(a_n- a_m)}
\end{equation}

The created complex networks are evaluated to obtain the PT for determination of the resilience of the electrical distribution system.

\subsection{Mathematical Model for energy sharing}
\label{subsection:Theoretical Results for the Coalitional Game}

Each house is indexed by $i \in \mathcal{N}= { 1, 2, . . . ,N }$. We consider that all houses in a microgrid have randomly invested in solar PV panels of area $a_i$ and energy storage units of capacity $B_i$ with ideal operating conditions. We consider that there are two fixed pricing periods in a day i.e., peak ($h$) and off-peak ($l$). The cost of obtaining energy from the grid ($\lambda$) is different during these two periods, with peak pricing $\lambda_h$ being higher than off-peak pricing $\lambda_l$. The energy consumption and generation of a household during the peak and off-peak periods are $H_{h_i}$, $H_{l_i}$ and $G_{h_i}$ and $G_{l_i}$ respectively. Under a net metering billing system, a household is reimbursed for the net electricity generated at a rate of $\mu$ at the end of a billing period. However, if the household consumes more power from the grid than it generates, it will be charged for the deficit power consumed at a rate of $\lambda$. The rate at which a household sells electricity back to the grid during peak and off-peak periods is $\mu_h$ and $\mu_l$ respectively. The pricing conditions are considered to be $\lambda_h \geq \mu_h$, $\lambda_l \geq \mu_l$, and $\mu_h \geq \lambda_l$. All houses will look to first utilize the available energy from both storage and solar units. The storage unit is completely discharged in the peak period and charged to its full capacity in the off-peak period. The daily cost of a household \cite{Victor} is given by:
\begin{multline}
    C(i)= \lambda_{h} (H_{h_i}-B_i-G_{h_i})^{+}-\mu_{h}(B_i+G_{h_i}-H_{h_i})^+ \\
    \qquad \qquad \qquad + \lambda_{l}(H_{l_i}+B_i-G_{l_i})^+  -\mu_{l}(G_{l_i}-H_{l_i}-B_i)^+
\end{multline}
where $(x)^+ = \max \{x,0\}$ for any real number $x$. All houses participating in energy sharing look to meet their joint consumption through aggregating their renewable energy resources. The coalitional game is defined as $G(\mathcal{N},C)$ with a finite number of prosumers from the set $\mathcal{N}$, each having value function $C$.  A coalition is any subset of prosumers $\mathcal{S} \subseteq \mathcal{N}$ where $\mathcal{N}$ is the grand coalition. The game $G(\mathcal{N},C)$ is 

\begin{enumerate}[label=\roman{*})]
    \item  subadditive i.e., it satisfies the condition $C(\mathcal{S})+C(\mathcal{T}) \ge C(\mathcal{S} \cup \mathcal{T})$ for a pair of coalitions $\mathcal{S},\mathcal{T} \subset \mathcal{N}$ which are disjoint, i.e., $\mathcal{S}\cap \mathcal{T} = \emptyset $,
    \item balanced i.e., it satisfies the condition $\sum_{\mathcal{S}\in 2^\mathcal{N}}^{}$ $ \alpha(S)$ $C(\mathcal{S})$ $= C(\mathcal{N})$ for any balanced map $\alpha : 2^\mathcal{N} \rightarrow [0,1] $
\end{enumerate}

Let $\xi_i$ denote the cost allocation for a prosumer $i \in \mathcal{S} $ and for a coalition $\mathcal{S}$, $\xi_\mathcal{S} = \underset{i\in \mathcal{S}}{\sum} \xi_i$. This allocation is said to be an imputation if it satisfies  $C(\mathcal{S}) =\xi_\mathcal{S}$ and $C(i) \ge \xi_i$ \cite{CHURKIN}. The set of all imputations is denoted by denoted by $\mathcal{I}$. The core, $\mathcal{C}$ of the game $G(\mathcal{N},C)$ is given by 
\begin{equation}
\mathcal{C} = \bigg(\xi \in \mathcal{I} : C(\mathcal{S}) \ge \xi(\mathcal{S}), \forall \mathcal{S} \in 2^\mathcal{N} \bigg)
\end{equation}
The aggregated peak-period and off-peak period house consumption and solar generation is given by $H_{h_\mathcal{N}} = \sum_{i\in \mathcal{N}}^{} H_{h_i}$ and $G_{h_\mathcal{N}} = \sum_{i\in \mathcal{N}}^{} G_{h_i}$, $H_{l_\mathcal{N}} = \sum_{i\in \mathcal{N}}^{} H_{l_i}$ and $G_{l_\mathcal{N}} = \sum_{i\in \mathcal{N}}^{} G_{l_i}$, respectively. The joint storage capacity is $B_\mathcal{S}=\sum_{i\in \mathcal{S}}^{} B_i$. The daily cost of a coalition $\mathcal{N}$ is given by
\begin{multline}
    C(\mathcal{N})= \lambda_{h}(H_{h_\mathcal{N}}-B_\mathcal{N}-G_{h_\mathcal{N}})^+
    -\mu_{h}(B_\mathcal{N}+G_\mathcal{N}-H_{h_\mathcal{N}})^+\\ 
    \qquad \qquad +\lambda_{l}(H_{l_\mathcal{N}}+B_\mathcal{N}-G_{l_\mathcal{N}})^+ -\mu_{l}(G_{l\mathcal{N}}-H_{l_\mathcal{N}}-B_\mathcal{N})^+
\end{multline}

The Bordareva-Shapley theorem states that if a coalitional game is balanced, it has a non-empty core. Our game is balanced, so it follows that the core is non-empty. This means that it is possible to find a cost allocation within the core of the coalition game. We have created a cost allocation {$\xi_i$} that is calculated using a simple analytical formula.

The cost allocation $\xi(i) :i \in \mathcal{N}$ and for a coalition, $\mathcal{S}$, $\xi(\mathcal{S}) = \underset{i\in \mathcal{S}}{\sum} \xi(i)$ satisfies (i) budget balance, i.e., $\underset{i\in \mathcal{N}}{\sum} \xi(i) = C(\mathcal{N})$ and (ii) individually rationality i.e., $\xi(i) \le C(i)$  for all $i \in N$. Therefore, they are mutations that belong to the core of the game and also satisfy the condition $\underset{i\in \mathcal{S}}{\sum} \xi_i \le C(\mathcal{S})$ for the coalition $\mathcal{S}\subseteq \mathcal{N}$. 

\begin{equation}
\resizebox{1\hsize}{!}{$
\xi(i)=\left\{
    \begin{aligned}
       &K_i \quad \textrm{if} \quad  H_{h_\mathcal{N}} \ge B_\mathcal{N}+G_{h_\mathcal{N}} \; \& \; H_{l_\mathcal{N}}+B_\mathcal{N} \ge G_{l_\mathcal{N}}\\ 
        &L_i \quad \textrm{if} \quad  H_{h_\mathcal{N}} < B_\mathcal{N}+G_{h_\mathcal{N}} \; \& \; H_{l_\mathcal{N}}+B_\mathcal{N} \ge G_{l_\mathcal{N}}\\ 
        &M_i \quad \textrm{if} \quad  H_{h_\mathcal{N}} \ge B_\mathcal{N} +G_{h_\mathcal{N}} \; \& \; H_{l_\mathcal{N}}+B_\mathcal{N} < G_{l_\mathcal{N}}\\ 
        &N_i \quad \textrm{if} \quad  H_{h_\mathcal{N}} < B_\mathcal{N}+G_{h_\mathcal{N}} \; \& \; H_{l_\mathcal{N}}+B_\mathcal{N} < G_{l_\mathcal{N}}&& 
    \end{aligned}\right.$}
\end{equation}
\noindent where,
\begin{flalign*}
 K_i&= \lambda_h (H_{h_i}-B_i-G_{h_i}) +\lambda_l (H_{l_i}+B_i-G_{l_i})\\
L_i&=-\mu_h (B_i+G_{h_i}-H_{h_i}) + \lambda_{l}(H_{l_i}+B_i-G_{l_i})\\
 M_i&= \lambda_{h} (H_{h_i}-B_i-G_{h_i}) -\mu_l(G_{l_i}-H_{l_i}-B_i)\\
 N_i&= -\mu_h (B_i+G_{h_i}-H_{h_i}) -\mu_l(G_{l_i}-H_{l_i}-B_i)&& 
\end{flalign*}

\subsection{Energy Prediction}
For the prediction of energy required from the grid, we use the ARIMA method for predicting time-series data. It incorporates autocorrelation measures within the time series data to forecast future values. The model's autoregression part measures a particular sample's dependency with a few past observations. These differences are integrated to make the data patterns stationary or minimize the obvious correlation with past data. The general form of the ARIMA $(p,d,q)$ model \cite{Book_stat} is,
\begin{equation}
\Phi_p(B) \nabla^d X_t=\Theta_q(B)\varepsilon_t
\end{equation}
where $\varepsilon_t$ is a random shock with mean zero and var ($\varepsilon_t$) = $\sigma^2_\varepsilon$. We indicate
lagged observations with the backshift operator $B$, defined to mean $BX_t=X_{t-1}$. The conventional notation for a time series is $X_t=1,2,...,n$. $\Phi_p(B)=\phi(B)=1-\phi_1B-...-\phi_pB^p$ is a $p^{th}$ degree polynomial and $\phi_p$ is the correlation with lag $p$. $\Theta_q(B)=\theta(B)=1-\theta_1B-...-\theta_q B^q$ is a $q^{th}$ degree polynomial and $\theta_q$ is the correlation with lag $q$. The differencing with order $d$ is represented by $\nabla^d$.

Determining the parameters $p$, $d$, and $q$ for implementing ARIMA models is crucial. We make use of the augmented-dickey fuller test (ADF), autocorrelation function (ACF), and Partial autocorrelation function (PACF) to determine the model parameters \cite{Vicor_AR}. The value of differencing $d$ is computed using ADF. Based on the values of PACF and ACF for different orders of differencing, the values of $p$ and $q$ are obtained, respectively. 

\begin{figure}[b]
  \centering
  \includegraphics[width=3.5in]{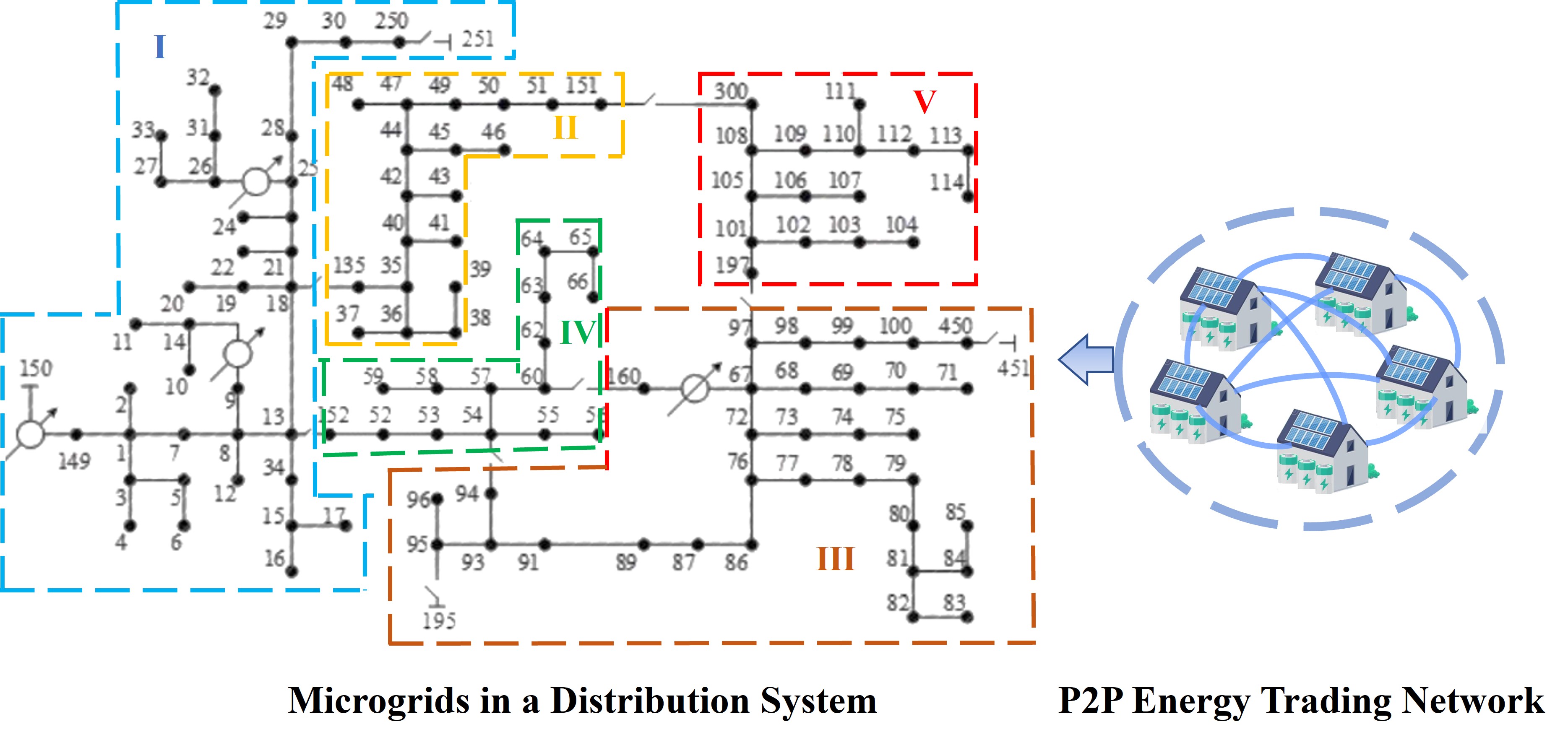}
  \caption{IEEE 123 bus test feeder partitioned into five microgrids with houses participating in P2P energy trading.}
  \label{fig:IEEE-123}
\end{figure}

\section{Simulation Study, Result Analysis and Discussion}
\label{section:Simulation}

\subsection{Data Processing}
The simulation model of the IEEE 123 test system is developed in GridLAB-D, an open-source software that provides a platform for distribution system design with the incorporation of renewable energy sources \cite{resiliency}. There are 11 switches in the test system that could be used to partition the system into the 5 possible microgrids (MG-I to MG-V) as shown in Fig. \ref{fig:IEEE-123}. Also, the system consists of 85 constant loads with a total real-power requirement of 3620.5 kW. We introduced a total of 516 houses at these constant loads with different floor areas of varying load profiles, solar PV panels with a size of approximately 10\% of floor area \cite{Victor}, and random storage capacity.  We simulated the model for 365 days with four-hour intervals considering the climate profile of Bakersfield, California \cite{Bakers}. From the simulation, we obtained the time-series data of active power, energy consumption, and generation for two cases, i.e., with and without renewable energy sources. This data is used for further analyzing the peer-to-peer trading between the houses in the IEEE 123 test system, as shown Fig. \ref{fig:IEEE-123}.

For the P2P trading scenario, we have considered a net-metering billing system with a time-of-use pricing scheme (ToU); peak and off-peak period price for buying (54\mbox{\textcentoldstyle}/kWh and 22\mbox{\textcentoldstyle}/kWh) and selling (30\mbox{\textcentoldstyle}/kWh and 13\mbox{\textcentoldstyle}/kWh) energy to and from the grid \cite{Victor-GM}. Prices for trading within the P2P network are given by the conditions for $\pi_h$ and $\pi_l$. We have considered the peak period from 8hrs to 20hrs and the off-peak period from 20hrs to 8hrs. In the peak period, the entire storage energy is utilized and in the off-peak period, the storage unit is charged to its total capacity for all houses. The solar energy is first utilized for the respective house, and then excess energy is either sold to the grid or P2P network for both periods. Using cooperative game theory, we have shown that all houses would rationally choose to connect to the P2P network and form a stable coalition with good cooperative behavior providing economic benefits for all houses. Table \ref{tab:data_ana} shows the average consumption, solar generation, and total storage capacity of all microgrid combinations.

\begin{table}[]
\setlength{\tabcolsep}{1pt}
\centering
\caption{Total Storage Capacity and Daily Average Consumption and Generation of all Microgrid Combinations}
\begin{tabular}{c c c c c}
    \toprule
    \multicolumn{1}{>{\centering}m{6em}}{Microgrid Combination} & \multicolumn{1}{>{\centering}m{4.5em}}{No. of Houses} & \multicolumn{1}{>{\centering}m{4.5em}}{Daily average consumption (kWh)} & \multicolumn{1}{>{\centering}m{4.5em}}{Daily average generation (kWh)} & \multicolumn{1}{>{\centering}m{4.5em}}{Total storage capacity (kWh)} \\
\midrule
\multicolumn{1}{c}{MG-I}        &113   &3341  & 2574  & 728  \\
\multicolumn{1}{c}{MG-II}      & 106     &3396  & 2455  & 612  \\
\multicolumn{1}{c}{MG-III}     & 161     &5234  & 3827  & 1010 \\
\multicolumn{1}{c}{MG-IV}      & 88     &2775  & 2025  & 1153 \\
\multicolumn{1}{c}{MG-V}        & 48    &1543  & 1153  & 742  \\
\midrule
\multicolumn{1}{c}{MG-I \& II}  & 219 & 6736  & 5029  & 1340 \\
\multicolumn{1}{c}{MG-I \& IV}  & 201 &6116  & 4599  & 2455 \\
\multicolumn{1}{c}{MG-II \& IV} & 194 & 6171  & 4480  & 2017 \\
\multicolumn{1}{c}{MG-II \& V}  & 154 & 4938  & 3609  & 1353 \\
\multicolumn{1}{c}{MG-III \& IV}  & 249 & 8010  & 5852  & 2415 \\
\multicolumn{1}{c}{MG-III \& V}  & 209  & 6777  & 4980  & 1752 \\
\midrule
\multicolumn{1}{c}{ALL MGs}      & 516   & 16289 & 12035 & 4497 \\
\bottomrule
\end{tabular}
\label{tab:data_ana}
\end{table}

\subsection{Result Analysis and Discussion}
\label{section:Results}

We now consider three scenarios; i) the houses in each microgrid participate in P2P trading, ii) the houses in two neighboring microgrids participate in P2P trading, and iii) all houses in all microgrids of the distribution system engage in P2P trading. The houses in each microgrid vary, and we analyze how P2P trading affects the cost, energy taken from the grid, and energy resilience of the system.

The cost analysis for all microgrid combinations for scenarios with and without P2P trading are tabulated in Table \ref{tab:MG cost}. We observe there are significant savings in cost due to P2P energy sharing for all cases. The time-series data of energy taken from the grid is used to create the visibility graph from which the PT values are computed. The energy taken from the grid and PT values are tabulated in Table \ref{tab:MG res}. It is observed that the PT values are improved when microgrids participate in P2P energy sharing. We observe in Table \ref{tab:PT comp}, the increase in cost savings and reduction in energy taken from the grid when microgrids collectively participate in P2P energy sharing. MG-III \& IV have the highest savings of \$4,864 and a reduction of 20,267kWh in energy requirement from the grid, and MG-I \& II have the lowest savings of \$243 and a reduction of 1,013kWh in energy requirement from the grid. The cost savings and energy requirement from the grid depends on the variations in the consumption patterns, solar generation patterns, and storage capacities for all the houses involved in P2P trading. 

\begin{table}[]
\setlength{\tabcolsep}{1pt}
\caption{Cost Analysis for each Microgrid and their Combinations for One Year}
\begin{tabular}{r c r r r r}
    \toprule
    \multicolumn{1}{>{\centering}m{6em}}{Microgrid Combination} & \multicolumn{1}{>{\centering}m{4.5em}}{No. of Houses} & \multicolumn{1}{>{\centering}m{4.5em}}{Without P2P (\$)} & \multicolumn{1}{>{\centering}m{4.5em}}{With P2P (\$)} & \multicolumn{1}{>{\centering}m{4.5em}}{Savings (\$)} & \multicolumn{1}{>{\centering}m{4.5em}}{\% Savings}\\
\midrule
\multicolumn{1}{c}{MG-I}           &113 & 40,969  & 37,288  & 3,681  & 8.98  \\
\multicolumn{1}{c}{MG-II}           & 106 & 64,732  & 60,790  & 3,942  & 6.09  \\
\multicolumn{1}{c}{MG-III}          & 161 & 93,330  & 86,211  & 7,119  & 7.63  \\
\multicolumn{1}{c}{MG-IV}           & 88  & 17,052  & 15,399  & 1,654  & 9.70  \\
\multicolumn{1}{c}{MG-V}            & 48  & 7,433   & 6,892   & 541   & 7.28  \\
\midrule
\multicolumn{1}{c}{MG-I \& II}   & 219 & 105,701 & 97,836  & 7,865  & 7.44  \\
\multicolumn{1}{c}{MG-I \& IV}   & 201 & 5,8021  & 50,431  & 7,591  & 13.08 \\
\multicolumn{1}{c}{MG-II \& IV} & 194 & 81,784  & 71,770  & 10,014 & 12.24 \\
\multicolumn{1}{c}{MG-II \& V}   & 154 & 72,165  & 64,439  & 7,726  & 10.71 \\
\multicolumn{1}{c}{MG-III \& IV} & 249 & 110,383 & 96,746  & 13,637 & 12.35 \\
\multicolumn{1}{c}{MG-III \& V}  & 209 & 100,763 & 89,873  & 10,890 & 10.81 \\
\midrule
\multicolumn{1}{c}{ALL MGs}         & 516 & 223,517 & 198,201 & 25,316 & 11.33 \\
\bottomrule
\end{tabular}
\label{tab:MG cost}
\end{table}

\begin{table}[]
\setlength{\tabcolsep}{2.5pt}
\caption{Percolation Threshold Values Computed from Energy taken from Grid for One Year for each Microgrid and their Combinations}
\begin{tabular}{>{\centering}m{5em} c r r r r}
    \toprule
    \multicolumn{1}{>{\centering}m{6em}}{Microgrid Combination} & \multicolumn{1}{>{\centering}m{3.2em}}{No. of Houses} & \multicolumn{1}{>{\centering}m{4.5em}}{Energy taken from grid without P2P (kWh)} & \multicolumn{1}{>{\centering}m{4.5em}}{Percolation Threshold without   P2P} & \multicolumn{1}{>{\centering}m{4.5em}}{Energy taken from grid  with P2P    (kWh)} & \multicolumn{1}{>{\centering}m{4.5em}}{Percolation Threshold with   P2P}\\
\midrule
\multicolumn{1}{c}{MG-I} & 113 & 668,618  & 0.12483 & 653,184  & 0.13379 \\
\multicolumn{1}{c}{MG-II} & 106 & 669,909  & 0.13087 & 653,369  & 0.13344 \\
\multicolumn{1}{c}{MG-III} & 161 & 1,048,131 & 0.13075 & 101,8137 & 0.13876 \\
\multicolumn{1}{c}{MG-IV} & 88  & 844,257  & 0.11549 & 837,368  & 0.12309 \\
\multicolumn{1}{c}{MG-V} & 48  & 450,203  & 0.11862 & 447,949  & 0.13584 \\
\midrule
\multicolumn{1}{c}{MG-I \& II} & 219 & 1,338,527 & 0.12909 & 1,305,540 & 0.14357 \\
\multicolumn{1}{c}{MG-I \& IV} & 201 & 1,512,875 & 0.11939 & 1,481,149 & 0.13818 \\
\multicolumn{1}{c}{MG-II \& IV} & 194 & 1,514,167 & 0.12692 & 1,472,324 & 0.14180 \\
\multicolumn{1}{c}{MG-II \& V} & 154 & 1,120,112 & 0.12244 & 1,087,804 & 0.14324
 \\
\multicolumn{1}{c}{MG-III \& IV} & 249 & 1,892,388 & 0.12591 & 1,835,237 & 0.14862
 \\
\multicolumn{1}{c}{MG-III \& V} & 209 & 1,498,334 & 0.11864 & 1,452,626 & 0.14518 \\
\midrule
\multicolumn{1}{c}{ALL MGs} & 516 & 3,681,119 & 0.12714 & 3,575,089 & 0.13994 \\
\bottomrule
\end{tabular}
\label{tab:MG res}
\end{table}

\begin{figure}[]
  \centering
  \includegraphics[width=3.5in]{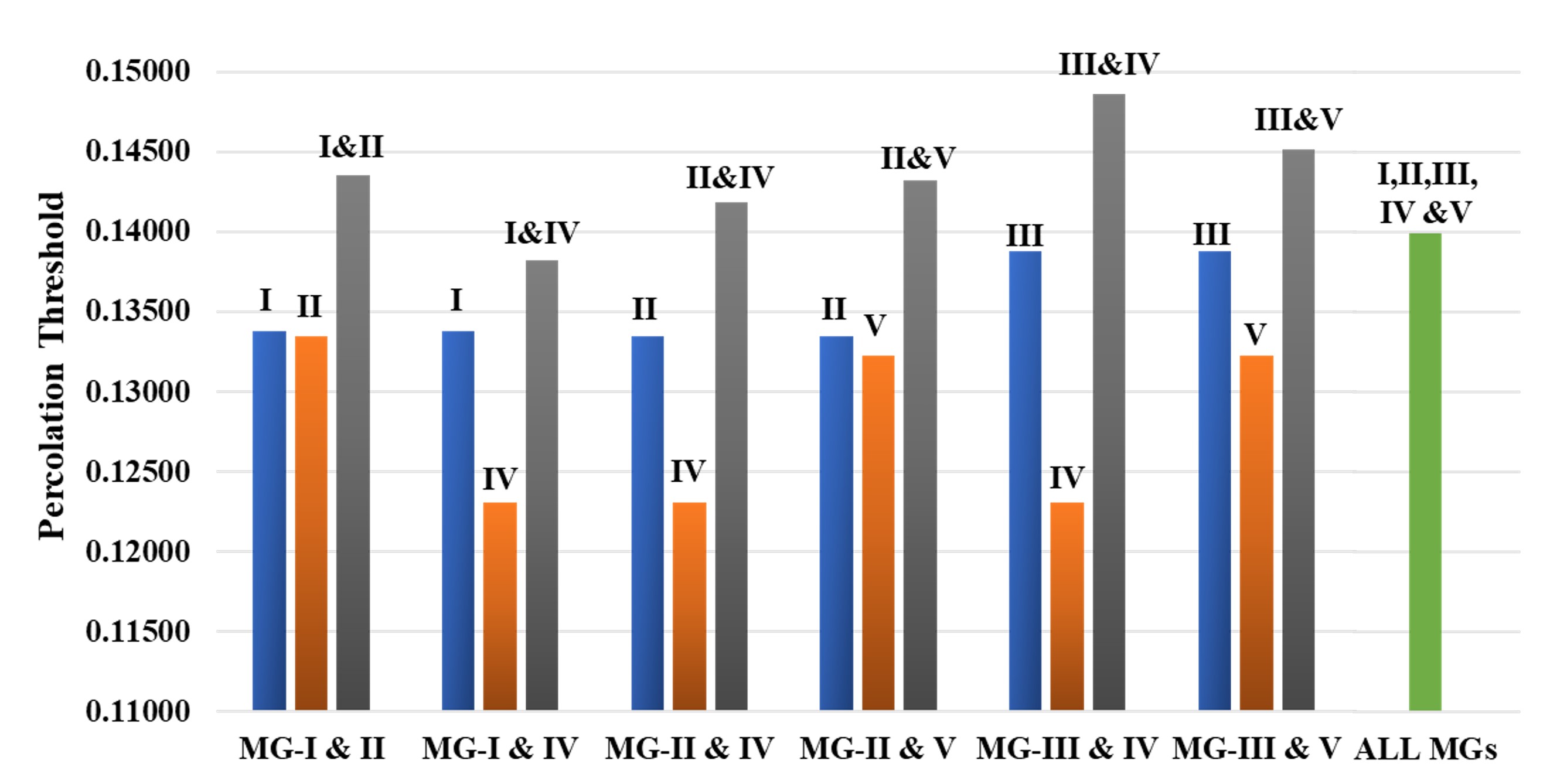}
  \caption{Comparison of percolation threshold values of all the considered cases.}
  \label{fig:PT_ana}
\end{figure}

In Fig. \ref{fig:PT_ana}, the PT values for combined microgrids with P2P are compared with the PT of individual microgrids. We observe that when two microgrids combine in P2P energy sharing, the combined PT is higher than all individual microgrid PT values. MG-I \& IV have the lowest combined PT value of 0.13818, and MG-III \& IV have the highest PT value of 0.14862. The PT values define the strength of the network; in our case, the grid dependency of a microgrid network and its ability to meet its energy requirements. PT values depend on the availability of green energy in the network and the number of interconnections of houses/nodes. From the results, it is observed that MG-III \& IV have the lowest cost savings and energy required from the grid when engaging in joint P2P, and their PT is the highest; indicating good network strength and energy resilience. All microgrids are joined together to engage in P2P trading and have a combined increase in savings of \$8,380 and a total reduction of 34,914kWh in energy requirement from the grid, with a combined PT value of  0.13994 which is higher than all the PT values of all individual microgrids.

\begin{table}[]
\setlength{\tabcolsep}{2pt}
\caption{Comparison of Energy taken from Grid and Cost Savings for One Year for Combined Microgrids and the sum of Individual Microgrid}
\begin{tabular}{>{\centering}m{5em} r r r r r r}
    \toprule
    \multicolumn{1}{>{\centering}m{6em}}{Microgrid Combination} & \multicolumn{1}{>{\centering}m{3.5em}}{a (kWh)} & \multicolumn{1}{>{\centering}m{3.5em}}{b (kWh)} & \multicolumn{1}{>{\centering}m{3.5em}}{c (kWh)} & \multicolumn{1}{>{\centering}m{3.5em}}{x (\$)} & \multicolumn{1}{>{\centering}m{3.5em}}{y (\$)} &
    \multicolumn{1}{>{\centering}m{3.5em}}{z (\$)}\\
    \midrule
\multicolumn{1}{c}{MG-I \& II} & 1,306,553 & 1,305,540 & 1,013  & 7,865  & 7,622  & 243  \\
\multicolumn{1}{c}{MG-I \& IV}   & 1,490,552 & 1,481,149 & 9,402  & 7,591  & 5,334  & 2,257 \\
\multicolumn{1}{c}{MG-II \& IV} & 1,490,737 & 1,472,324 & 18,413 & 10,014 & 5,595  & 4,419 \\
\multicolumn{1}{c}{MG-II \& V}   & 1,101,318 & 1,087,804 & 13,513 & 7,726  & 4,483  & 3,243 \\
\multicolumn{1}{c}{MG-III \& IV} & 1,855,504 & 1,835,237 & 20,267 & 13,637 & 8,773  & 4,864 \\
\multicolumn{1}{c}{MG-III \& V}  & 1,466,085 & 1,452,626 & 13,460 & 10,890 & 7,660  & 3,230 \\
\midrule
\multicolumn{1}{c}{ALL MGs} & 3,610,006 & 3,575,089 & 34,917 & 25,316 & 16,936 & 8,380 \\
\bottomrule
\end{tabular}\\

a - Sum of individual energy taken from the grid for each MG with P2P,\\
b - Combined energy is taken from the grid by two MGs with P2P,\\
c - Energy reduction (a-b),\\
x - Cost savings of two combined microgrids with P2P,\\
y - Sum of individual cost savings with P2P for each microgrid, \\
z - Increase in cost savings (x-y).
\label{tab:PT comp}
\end{table}

For all cases, we have considered that there exists a P2P network between all the houses participating in the energy sharing. Also, there may be transmission loss for energy sharing between households located at long distances and some costs for maintaining the P2P network. Any additional loss and costs could be shared equally by all participants in the energy sharing. This would not affect the fundamental game model and analysis. The major focus of our work is to show that trading of excess renewable energy for more number of houses would result in a significant reduction in energy requirement from the grid leading to a more resilient system and also providing cost benefits.

\begin{figure}[]
  \centering
  \includegraphics[width=3in]{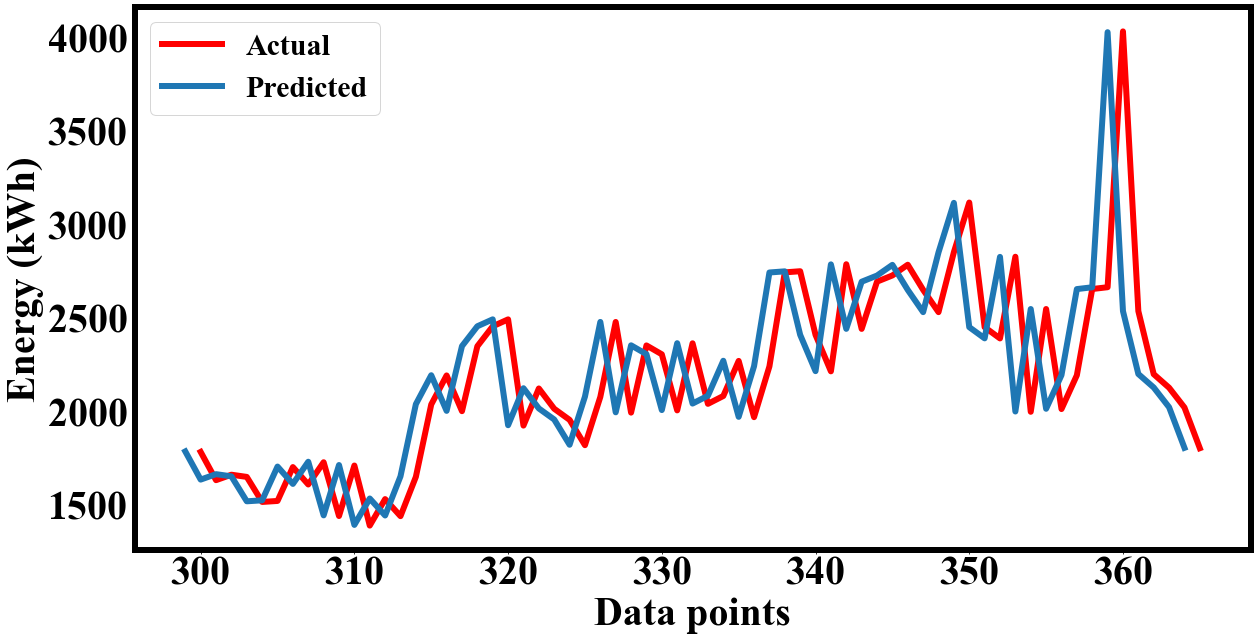}
  \caption{Comparison of the predicted data with actual data for validation.}
  \label{fig:prediction_comp}
\end{figure}

The prediction of energy taken from the grid is crucial to understand the resilience of a microgrid. Analyzing consumption patterns, capacity planning, and load management helps in the improvement of emergency preparedness leading to greater efficiency, reduced energy waste, and enhanced resilience during times of crisis. The time-series data is for one year with 365 data points. For predicting the energy taken from the grid, we perform the simulation to obtain the energy data of the system for all 365 data points. We use the first 300 data points to obtain the parameters of ARIMA and use the remaining 65 data points, i.e., 65 days of data, to analyze the prediction. Using the 300 data points, the parameters of $(p,q,d)$ are found to be $(0,2,1)$. The probability distribution of the residual errors was found to be a Gaussian distribution, which is an important check when using the ARIMA model. The prediction of energy taken from the grid for the last 65 days is compared with the actual values as shown in Fig. \ref{fig:prediction_comp}. The prediction is validated through the test scores; the R-squared (R\textsuperscript{2}) value is 0.9999, which is close to 1 (maximum), Root Mean Square Error (RMSE) is 1.8500 and Mean Absolute Error (MAE) is 1.4854. We have computed the percolation value for the last 65 days based on the prediction and found it to be 0.13706.

\section{Conclusion}
\label{section:Conclusion}

We have analyzed in detail how energy sharing benefits microgrids in terms of cost, energy consumption from the grid, and energy resilience. Possible combinations for microgrid formation were identified in a standard IEEE 123 test bus feeder by considering the switching operation. The cooperative game theory approach was used to model energy sharing in a peer-to-peer network that allows all houses to obtain cost benefits by either buying from or selling to the P2P network. To measure the energy resilience of a system, we computed PT values for the total energy taken from the grid when the microgrid participates in P2P trading. We analyzed the benefits of P2P within a microgrid, with neighboring microgrids and all microgrids combined together in a distribution system. The results show significant improvement in the PT values, reduced energy taken from the grid, and increased cost savings for all microgrid combinations. Additionally, we introduced the prediction of future utility grid energy requirements of a microgrid. This allows us to evaluate the resilience and help in the planning and operation of a microgrid.

\bibliographystyle{IEEEtran}
\bibliography{main}

\end{document}